# Neural Network Attribution Methods for Problems in Geoscience: A Novel Synthetic Benchmark Dataset

by

Antonios Mamalakis[1*], Imme Ebert-Uphoff[2,3], and Elizabeth A. Barnes[1]

[1]Department of Atmospheric Science, Colorado State University, Fort Collins, CO, USA
[2]Department of Electrical and Computer Engineering, Colorado State University, Fort Collins, CO, USA
[3]Cooperative Institute for Research in the Atmosphere, Colorado State University, Fort Collins, CO, USA

*Corresponding author: Antonios Mamalakis (amamalak@rams.colostate.edu)

**Abstract.** Despite the increasingly successful application of neural networks to many problems in the geosciences, their complex and nonlinear structure makes the interpretation of their predictions difficult, which limits model trust and does not allow scientists to gain physical insights about the problem at hand. Many different methods have been introduced in the emerging field of eXplainable Artificial Intelligence (XAI), which aim at attributing the network's prediction to specific features in the input domain. XAI methods are usually assessed by using benchmark datasets (like MNIST or ImageNet for image classification). However, an objective, theoretically-derived ground truth for the attribution is lacking for most of these datasets, making the assessment of XAI in many cases subjective. Also, benchmark datasets specifically designed for problems in geosciences are rare. Here, we provide a framework, based on the use of additively separable functions, to generate attribution benchmark datasets for regression problems for which the ground truth of the attribution is known *a priori*. We generate a large benchmark dataset and train a fully-connected network to learn the underlying function that was used for simulation. We then compare estimated heatmaps from different XAI methods to the ground truth in order to identify examples where specific XAI methods perform well or poorly. We believe that attribution benchmarks as the ones introduced herein are of great importance for further application of neural networks in the geosciences, and for more objective assessment and accurate implementation of XAI methods, which will increase model trust and assist in discovering new science.

**Impact Statement.** The fidelity of methods of eXplainable Artificial Intelligence (XAI) is difficult to assess and often done subjectively, since there is no ground truth about how the explanation should look. Here, we introduce a general approach to create synthetic problems, where the ground truth of the explanation is a priori known, thus, allowing for objective XAI assessment. We generate a synthetic climate prediction problem, and we test popular XAI methods in explaining the predictions of a dense neural network. It is shown that systematic strengths and weaknesses can be easily identified, which have been overlooked in other applications. Our work highlights the importance of introducing objectivity in the assessment of XAI to increase model trust and assist in discovering new science.

## 1. Introduction

Within the last decade, neural networks (NNs; LeCun et al., 2015) have seen tremendous application to the field of geosciences (Lary et al., 2016; Shen, 2018; Karpatne et al., 2018; Reichstein et al., 2019; Bergen et al., 2019; Barnes et al., 2019, Rolnick et al., 2019; Ham et al., 2019; Sit et al., 2020), owing in part to their impressive performance in capturing nonlinear system behavior (LeCun et al., 2015), and the increasing availability of observational and simulated data (Overpeck et al., 2011; Guo, 2017; Agapiou, 2017; Reinsel et al., 2018). However, due to their complex structure, NNs are difficult to interpret (the so-called "black box" model). This limits their reliability and applicability since scientists cannot verify when a prediction is successful for the right reasons (i.e., they cannot test against "clever Hans" prediction models; Lapuschkin et al., 2019) or improve the design of a model that is performing poorly (see for example Ebert-Uphoff and Hilburn, 2020). Also, when applying NNs to new problems, the interpretability problem does not allow scientists to gain physical insights about the connections between the input variables and the prediction, and generally about the problem at hand. To address the interpretability problem, many different methods have been developed in recent years (Zeiler and Fergus, 2013; Springenberg et al., 2015; Bach et al., 2015; Shrikumar et al., 2016; 2017; Smilkov et al., 2017; Kindermans et al., 2017a; Sundararajan et al., 2017; Montavon et al., 2017; Ancona et al., 2019) in the emerging field of post hoc eXplainable Artificial Intelligence (XAI; Buhrmester et al., 2019; Tjoa and Guan, 2019; Das and Rad, 2020). These methods aim at a post hoc explanation of the prediction of a NN by determining its attribution or sensitivity to specific features in the input domain (usually referred to as attribution heatmaps or saliency maps), thus highlighting relationships that may be interpreted physically, and making the "black box" more transparent (McGovern et al., 2019). Given that physical understanding is highly desirable to accompany any successful model in the geosciences, XAI methods are expected to be a real game-changer for further application of NNs in this field (Toms et al., 2020; Barnes et al., 2020).

Despite their high potential, many XAI methods have been shown to not honor desirable properties (e.g., "completeness" or "implementation invariance"; see Sundararajan et al., 2017), and in general, face nontrivial limitations for specific problem setups (Ancona et al., 2018; Kindermans et al., 2017b; Rudin, 2019; Dombrowski et al., 2020). Thus, thorough investigation and assessment of XAI methods is of vital importance to be reliably applied in new scientific problems. So far, the assessment of the outputs of different XAI methods in geoscientific research (and in computer science) has been mainly based on applying these methods to benchmark problems, where the scientist is expected to know what the heatmaps should look like, thus being able to judge the performance of the XAI method in question. Examples of benchmark problems for the geosciences include the classification of El Niño or La Niña years or seasonal prediction of regional hydroclimate (Ham et al., 2019; Toms et al., 2020). In computer science, commonly used benchmark datasets for image classification problems include MNIST or ImageNet among others (LeCun et al., 1998; Russakovsky et al., 2015). A second way to assess the output of an XAI method is through deletion/insertion techniques (Samek et al., 2017; Petsiuk et al., 2018; Qi et al., 2020), where highlighted features are deleted from the full image (or added to a grey image). If the XAI method has highlighted important features for the prediction, then the performance of the network is expected to decrease (improve) as these features are being deleted (added).

Although the above are good ways to gain insight about the performance of different XAI methods, in both cases, a ground truth of attribution is lacking, limiting the degree to which one can objectively assess their fidelity. When using standard benchmark datasets, the scientist typically assesses the XAI methods based on visual inspection of the results and their prior knowledge and understanding of the problem at hand. However, the human perception of an explanation is subjective and can often be biased. Thus, human perception alone is not a solid criterion for assessing trustworthiness. For example, although it might make sense to a human that an XAI method highlights the ears or the nose of a cat for an image successfully classified as "cat", it is not proof that these are the features the network actually based its decision on. The relative importance of these (and other) features to the prediction is always task- and/or dataset-dependent, and since no ground truth of attribution is provided for each considered dataset, the scientist can only subjectively (but not objectively) assess the performance of the XAI methods. Similarly, when using the deletion (or the insertion) technique, although the scientist can assess which XAI method highlights more important features relatively to other XAI methods (i.e., this would be the method that corresponds to the most abrupt drop in performance as the highlighted features are being deleted), there is no proof that these features are indeed the most important ones; it could be the case that even the best performing XAI method corresponds to a much less abrupt performance drop than the most abrupt possible. The lack of objectivity in the assessment of XAI as described above involves high risks of cherry-picking specific samples/methods and reinforcing individual biases; Leavitt and Morcos (2020). Moreover, we note that benchmark datasets that refer to regression problems are very rare, which is problematic, since many geoscientific applications are better approached as regression rather than classification problems.

Given the above and with the aim of a more falsifiable XAI research (Leavitt and Morcos, 2020), in this paper we provide, for the first time, a framework to generate nonlinear benchmark datasets for geoscientific problems and beyond, where the importance of each input feature to the prediction is objectively derivable and known *a priori*. This *a priori* known attribution for each sample can be used as ground truth for evaluating different XAI methods and identifying examples where they perform well or poorly. We refer to such synthetic datasets as "attribution benchmark datasets", to distinguish from benchmarks where no ground truth of the attribution is available. Our framework is outlined here for regression problems (but can be extended into classification problems too), where the input is a 2D field (i.e., a single-channel image); commonly found in geoscientific applications (e.g., DelSole and Banerjee, 2017; Ham et al., 2019; Toms et al., 2020; Stevens et al., 2021).

We describe our synthetic framework and generate an attribution benchmark in section 2. Next we train a fully-connected NN on the synthetic dataset and apply different XAI methods to explain the network (section 3). We compare the estimated heatmaps with the ground truth in order to thoroughly and objectively assess the performance of the XAI methods considered here (section 4). In section 5, we state our conclusions.

## 2. A nonlinear attribution benchmark dataset

Let us consider the illustrative problem of predicting regional climate from global 2D fields of sea surface temperature (SST; see e.g., DelSole and Banerjee, 2017; Stevens et al., 2021). The general idea of this paper is summarized in Fig. 1. We start by generating $N$ realizations of an input random vector $\mathbf{X} \in \mathbb{R}^d$ (e.g., $N$ synthetic samples of a vectorized 2D SST field). We use a nonlinear function $F: \mathbb{R}^d \to \mathbb{R}$, which represents the physical system of our problem setting (e.g., the climate system), to map each realization $\mathbf{x}_n$ into a scalar $y_n$, and generate the output random variable Y (e.g., regional climatic variable). We then train a fully-connected NN to approximate the underlying function $F$ and compare the XAI heatmaps estimated by different XAI methods with the ground truth of attribution derived from $F$ in order to thoroughly and objectively assess their performance.

In this section, we describe how to generate synthetic datasets of the input $\mathbf{X}$ and the output Y. Although here we present results for a climate prediction setting for illustration, our framework is generic and applicable to a large number of problem settings in the geosciences and beyond. Regarding the adopted network architecture, we note that a fully-connected NN is explored here as a first step and to illustrate our framework since this type of architecture has been used in many recent climate-XAI studies (e.g., Toms et al., 2020; Barnes et al., 2019; 2020), but many other network architectures could be used as well (e.g., convolutional networks; see Mamalakis et al., 2022).

### 2.1. Input variables

We start by randomly generating $N = 10^6$ independent realizations of an input vector $\mathbf{X} \in \mathbb{R}^d$. Although arbitrary, the distributional choice of the input is decided with the aim of being a reasonable proxy of the independent variable of the physical problem of interest. Here, the input series represent monthly global fields of SST anomalies (deviations from the seasonal cycle) at a $10^o \times 10^o$ resolution (fields of $d = 458$ variables; see step 1 in Fig. 1). We generate the SST anomaly fields from a multivariate Normal Distribution $MVN(\mathbf{0}, \mathbf{\Sigma})$, where $\mathbf{\Sigma}$ is the covariance matrix and represents the dependence between SST anomalies in different grid points (or pixels in image classification settings) around the globe (spatial dependence). The matrix $\mathbf{\Sigma}$ is set equal to the sample correlation matrix that is estimated from monthly SST observations[1]. If a user wants to eliminate spatial dependence, then a good choice might be $\mathbf{\Sigma} = \sigma^2 I_d$, where $\sigma^2$ is the variance and $I_d$ is the identity matrix. We note that we decided to generate a large amount of data $N = 10^6$ (much larger than what is usually available in reality), so that we can achieve a near perfect NN prediction accuracy, and make sure that the trained network (labeled as $\hat{F}$) approximates very closely the underlying function $F$. Only under this condition, is it fair to use the derived ground truth of attribution as a benchmark for the XAI methods. Yet, we highlight that discrepancies between the two shall always exist to a certain degree due to $\hat{F}$ not being identical to $F$.

---

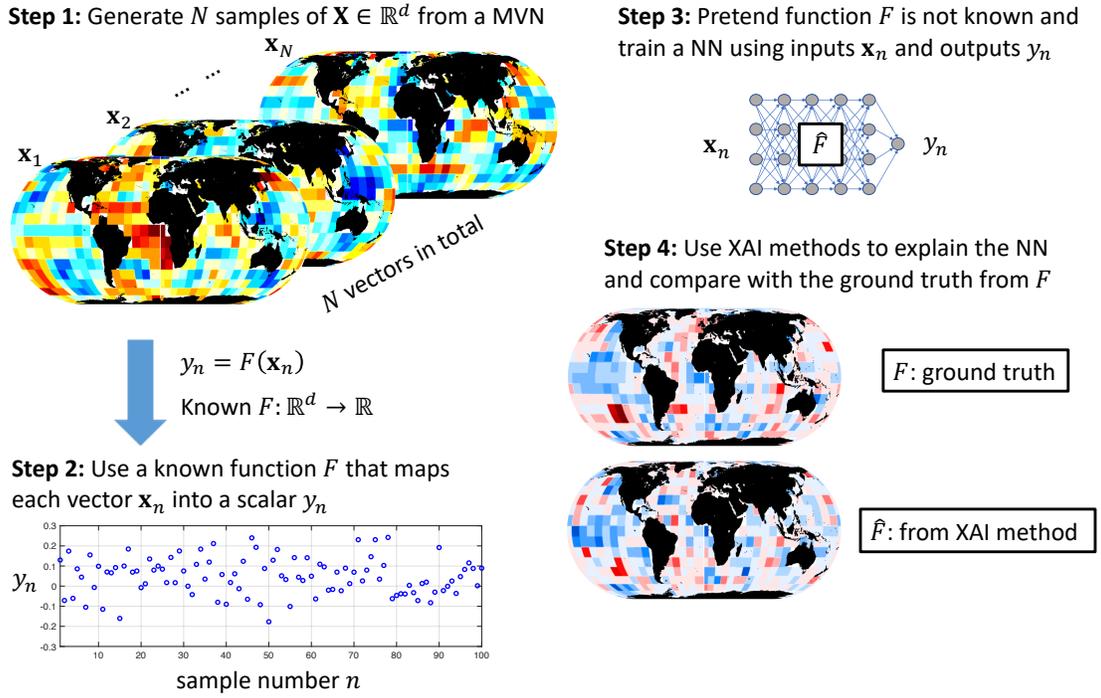

Fig. 1. Schematic overview of the general idea of the paper for a climate prediction setting. In step 1 (section 2), we generate $N$ independent realizations of a random vector $\mathbf{X} \in \mathbb{R}^d$ from a multivariate Normal Distribution. In step 2 (also section 2), we generate a response $Y \in \mathbb{R}$ to the synthetic input $\mathbf{X}$, using a known nonlinear function $F$. In step 3 (section 3), we train a fully-connected NN using the synthetic data $\mathbf{X}$ and Y to approximate the function $F$. The NN learns a function $\hat{F}$. Lastly, in step 4 (section 4), we compare the XAI heatmaps estimated from different XAI methods to the ground truth, which represents the function $F$ and has been objectively derived for any sample $n = 1, 2, \ldots, N$.

## 2.2. Synthetic response based on additively separable functions

We next create a nonlinear response $Y \in \mathbb{R}$ to the synthetic input $\mathbf{X} \in \mathbb{R}^d$ (see step 2 in Fig. 1), using a real function $F: \mathbb{R}^d \to \mathbb{R}$. For any sample $n = 1, 2, \ldots, N$, the response of our system $y_n$ to the input $\mathbf{x}_n$ is given as $y_n = F(\mathbf{x}_n)$ or after dropping the index $n$ for simplicity and relating the random variables instead of the samples, $Y = F(\mathbf{X})$.

Theoretically, the function $F$ can be any function the user is interested to benchmark an NN against. However, in order for the final synthetic dataset to be useful as an attribution benchmark for XAI methods, function $F$ needs to: i) have such a form so that the attribution of each of the responses Y to the input variables is objectively derivable, and ii) be nonlinear, as is the case in the majority of applications in geoscience.

The simplest form for $F$ so that the above two conditions are honored is when $F$ is an *additively separable function*, i.e. there exist local functions $C_i$, with $i = 1, 2, \ldots, d$, so that:

$$F(\mathbf{X}) = F(X_1, X_2, \ldots, X_d) = C_1(X_1) + C_2(X_2) + \cdots + C_d(X_d)$$

where, $X_i$ is the random variable at grid point $i$. Under this setting, the response Y is the sum of local responses at grid points $i$, and although the local functions $C_i$ may be independent from each other, one can also apply functional dependence by enforcing neighboring $C_i$ to behave similarly, when the physical problem of interest requires it (see next subsection). Moreover, as long as the local functions $C_i$ are nonlinear, so is the response Y, which satisfies our aim. Most importantly, with $F$ being an additively separable function, and considering a zero baseline, the contribution of each of the variables $X_i$ to the response $y_n$ for any sample $n$, is by definition equal to the value of the corresponding local function, i.e., $R_{i,n}^{true} = C_i(x_{i,n})$. This satisfies the basic desirable property that any response can be objectively attributed to the input.

Of course, this simple form of $F$ comes with the pitfall that it may not be exactly representative of some more complex physical problems. However, for this study we are not trying to capture all possible functional forms of $F$, but rather, provide a sample form of $F$ that honors the two desirable properties for benchmarking XAI methods and is complex enough to be considered representative of climate prediction settings (see discussion in the following section). We also note that by changing the form of local functions $C_i$ (and the dimension $d$), one can create theoretically infinite different forms of $F$ and of attribution benchmarks to test XAI methods against. Next, we define the local functions $C_i$ that we use in this study.

## 2.3. Local functions

The simplest form of the local functions is linear, i.e., $C_i(X_i) = \beta_i X_i$. In this case, the response Y falls back to a traditional linear regression response, which is not necessarily very interesting (there is no need to train a NN to describe a linear system), and it is certainly not representative of the majority of geoscience applications. More interesting responses to benchmark a NN or XAI methods against are responses where the local functions have a nonlinear form, e.g., $C_i(X_i) = \sin(\beta_i X_i + \beta_i^0)$, or $C_i(X_i) = \beta_i X_i^2$, etc.

In this study, to avoid prescribing the form of nonlinearity, we defined the local functions to be piece-wise linear (PWL) functions, with number of break points $K$, and with the condition $C_i(0) = 0$, for any grid point $i = 1, 2, \dots, d$. Our inspiration for using PWL functions is the use of ReLU (a piece-wise linear function with $K = 1$) as the activation function in NN architectures: indeed, a piece-wise linear response can describe highly nonlinear behavior of any form as the value of $K$ increases. Regarding the suitability of this choice to represent climate data, the condition $C_i(0) = 0$ leads to a reasonable condition for climate applications, that $y|_{x=0} = F(\mathbf{0}) = 0$, i.e., if the SST is equal to the climatological average, then the response Y is also equal to the climatological average. Moreover, the use of PWL functions allows us to model asymmetric responses of the synthetic system to the local inputs, which is frequently met in real climate prediction settings. For example, it is well established in the climate literature that the response of the extratropical hydroclimate to the El Niño-Southern Oscillation (ENSO) is not linear, and the effect of El Niño and La Niña events on the extratropics is not symmetric (Zhang et al., 2014; Feng et al., 2017). Lastly, we have used composite analysis and found that the generated Y series exhibit an ENSO-like dependence regarding extreme samples, with the highest 10% of $y$ values corresponding to a La Niña-like pattern and the lowest 10% of $y$ values corresponding to an El Niño-like pattern (not shown). This provides more evidence that our generated dataset honors both the spatial patterns of observed SSTs in the synthetic input (due to the use of the observed spatial correlation in the generation process) and the importance of ENSO variability for extreme events as manifested in the relation between the synthetic input and output. So, in this study, we will generate the Y series using PWL local functions, but we note that the benchmarking of XAI methods can also be performed using other types of additively separable responses.

A schematic example of a local piece-wise linear function $C_i$, for $K = 5$ that is used herein, is presented in Fig. 2. For each grid point $i$, the break points $l_k$, $k = 1, 2, \dots, K$ are obtained as the empirical quantiles of the synthetic series of $X_i$ that correspond to probability levels chosen randomly from a uniform distribution (also, note that we enforce that the point $x = 0$ is always a break point). The corresponding slopes $\beta_i^1, \beta_i^2, \dots, \beta_i^{K+1}$ are chosen randomly by generating $K + 1$ realizations from a $\text{MVN}(\mathbf{0}, \mathbf{\Sigma})$, where $\mathbf{\Sigma}$ is again estimated from SST observations and is used to enforce spatial dependence in the local functions. In Fig. 2, the map of $\beta_i^6$ (the slope for $X_i \in (l_5, \infty)$) is shown for all grid points in the globe, and the local functions $C_i$ for three points A, B, and C are also presented. First, the spatial coherence of the slopes $\beta_i^6$ is evident, with positive slopes over e.g., most of the northern Pacific and Indian Ocean and negative slopes over e.g., the eastern tropical Pacific. Second, the local functions at the neighboring points A and B are very similar to each other, consistent with the functional spatial dependence that we have specified. Lastly, the local function at point C very closely approximates a linear function. Indeed, in the way that the slopes $\beta_i^1, \beta_i^2, \dots, \beta_i^{K+1}$ are randomly chosen, it is possible that the local functions at some grid points end up being approximately linear. However, based on Monte Carlo simulations, we have established that the higher the value of $K$, the more unlikely it is to obtain approximately linear local functions (not shown).

Before moving forward, we wish to again highlight that although the total response Y is nonlinear and potentially very complex, the contributions of the input variables to the response are known and equal to the corresponding local functions (for a zero baseline), simply because the function $F$ is an additively separable function.

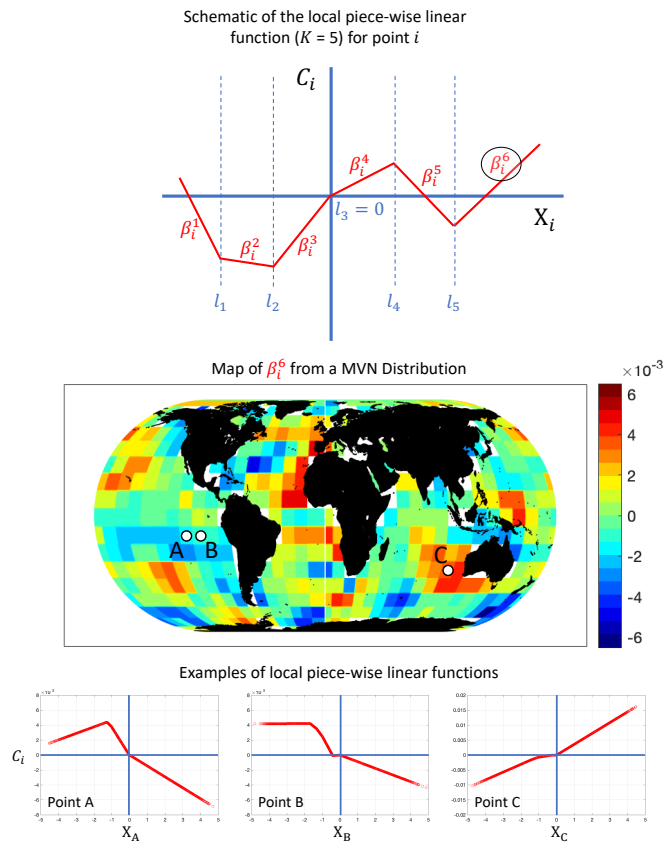

Fig. 2. Schematic representation and actual examples of local piece-wise linear functions $C_i$, for $K = 5$.

## 3. Neural network architecture and XAI methods

So far, we generated $N = 10^6$ independent realizations of an input vector $\mathbf{X} \in \mathbb{R}^d$ (with $d = 458$) and of an output scalar response $Y \in \mathbb{R}$, using an additively separable function $F$, with PWL local functions and $K = 5$. Next, we train a fully-connected NN to learn the function $F$ (see step 3 in Fig. 1), using the first 900,000 samples for training and the last 100,000 samples for testing. Apart from assessing the prediction performance, the testing samples will also be used to assess the performance of different post hoc, local XAI methods that have been commonly used in the literature and that are defined below.

### 3.1. Neural network

To approximate the function $F$, we used a fully-connected NN (with ReLU activation functions), with six hidden layers, each one containing 512, 256, 128, 64, 32, and 16 neurons, respectively. The output layer of our network contains a single neuron, which acts as the network's prediction and uses a "linear" activation function (also known as "identity" or "no activation"). We used the mean squared error as our loss function for training, and given that our synthetic input follows a multivariate normal distribution with zero mean vector and unit variances (and specific dependence structure), we did not need to apply any standardization/preprocessing of the data.

We do not argue that this is the optimum architecture to approach the problem since this is not the focus of our study. Instead, what is important is to achieve high enough performance, so that the benchmarking of XAI methods is as objective and fair as possible. The coefficient of determination of the NN prediction in the testing sample was slightly higher than $R^2 = 99\%$, which suggests that the NN can explain 99% of the variance in Y. As a benchmark to the NN, we also trained a linear regression model. The performance of the linear model was much poorer, with $R^2 = 65\%$ for the testing data.

### 3.2. XAI methods

For our analysis, we consider local, post hoc XAI methods that have commonly been used in the field of geoscience (e.g., McGovern et al., 2019; Toms et al., 2020; Barnes et al., 2019; 2020; Ebert-Uphoff and Hilburn, 2020).

**1) Gradient:** This method is among the simplest (conceptually) and very commonly used methods to explain an NN prediction. In this method, one calculates the partial derivative of the network's output with respect to each of the input variables $X_i$, at the specific sample $n$ in question:

$$R_{i,n} = \frac{\partial \hat{F}}{\partial X_i}\bigg|_{X_i = x_{i,n}} \quad (2)$$

where $\hat{F}$ is the function learned by the NN, as an approximation to the function $F$. This method estimates the *sensitivity* of the network's output to the input variable $X_i$. The motivation for using the Gradient method is that if changing the value, $x_{i,n}$, of a grid point in a given input sample is shown to cause a large difference in the NN output value, then that grid point might be important for the prediction. Furthermore, calculation of the Gradient is very convenient, as it is readily available in any neural network training environment, contributing to the method's popularity. However, as we will see in the next section 4, the sensitivity of the prediction of the network to the input is conceptually different from its attribution.

**2) Smooth Gradient:** This sensitivity method was introduced in Smilkov et al. (2017) and is very similar to the method Gradient, except that it aims to obtain a more robust estimation of the local partial derivative by averaging the gradients over a perturbed number of inputs with added noise:

$$R_{i,n} = \frac{1}{m} \sum_{j=1}^{m} \frac{\partial \hat{F}}{\partial X_i}\bigg|_{X_i = x_{i,n} + e_{i,n,j}} \quad (3)$$

where $m$ is the number of perturbations, and $e_{i,n,j}$ comes from a standard Normal Distribution.

**3) Input*Gradient:** As is evident from its name, in this method (Shrikumar et al., 2016; 2017) one multiplies the local gradient with the input itself:

$$R_{i,n} = x_{i,n} * \frac{\partial \hat{F}}{\partial X_i}\bigg|_{X_i = x_{i,n}} \quad (4)$$

In contrast to the previous two, this method quantifies the *attribution* of the output to the input. Attribution methods aim to quantify the marginal contribution of each feature to the output value (a different objective of explanation from sensitivity). The Input*Gradient method is used in the majority of XAI studies due to its conceptual simplicity.

**4) Integrated Gradients:** This method (Sundararajan et al., 2017) is also an attribution method similar to the Input*Gradient method, but aims to account for the fact that in nonlinear problems the derivative is not constant, and thus, the product of the local derivative with the input might not be a good approximation of the input's contribution. This method considers a reference (baseline) vector $\hat{x}$ (e.g., for which the network's output is zero, i.e., $\hat{F}(\hat{x}) = 0$), and the attribution is equal to the product of the distance of the input from the reference point with the average of the gradients at points along the straightline path from the reference point to the input:

$$R_{i,n} = (x_{i,n} - \hat{x}_i) * \frac{1}{m} \sum_{j=1}^{m} \frac{\partial \hat{F}}{\partial X_i}\bigg|_{X_i = \hat{x}_i + \frac{j}{m}(x_{i,n} - \hat{x}_i)} \quad (5)$$

where $m$ is the number of steps in the Riemann approximation.

Next, we consider different implementation rules of the attribution method Layer-wise Relevance Propagation (LRP; Bach et al., 2015; Samek et al., 2016). In the LRP method, one sequentially propagates the prediction $\hat{F}(\mathbf{x}_n)$ back to neurons of lower layers, obtaining the intermediate contributions to the prediction for all neurons until the input layer is reached and the attribution of the prediction to the input $R_{i,n}$ is calculated.

**5) LRP$_z$:** In the first LRP rule we consider, the back propagation is performed as follows:

$$R_i^{(l)} = \sum_j \frac{z_{ij}}{z_j} R_j^{(l+1)} \quad (6)$$

where $R_j^{(l+1)}$ is the contribution of the neuron $j$ at the upper layer $(l+1)$, and $R_i^{(l)}$ is the contribution of the neuron $i$ at the lower layer $(l)$. The propagation is based on the ratio of the localized preactivations $z_{ij} = w_{ij} x_i$ during prediction time and their respective aggregation $z_j = \sum_i z_{ij} + b_j$. Because this rule might lead to unbounded contributions as $z_j$ approaches zero (Bach et al., 2015), additional advancements have been proposed.

**6) LRP$_{\alpha\beta}$:** In this rule (Bach et al., 2015), positive and negative preactivations $z_{ij}$ are considered separately, so that the denominators are always nonzero:

$$R_i^{(l)} = \sum_j \left( \alpha \frac{z_{ij}^+}{z_j^+} + \beta \frac{z_{ij}^-}{z_j^-} \right) R_j^{(l+1)} \quad (7)$$

where

$$z_{ij}^+ = \begin{cases} z_{ij}; & z_{ij} > 0 \\ 0 \end{cases} \qquad z_{ij}^- = \begin{cases} 0 \\ z_{ij}; & z_{ij} < 0 \end{cases}$$

In our study, we use the commonly used rule with $\alpha = 1$ and $\beta = 0$, which considers only positive preactivations (Bach et al., 2015).

**7) Deep Taylor Decomposition:** For each neuron $j$ at an upper layer $(l+1)$, this method (Montavon et al., 2017) computes a rootpoint $\hat{x}_i^j$ close to the input $x_i$, for which the neuron's contribution is zero, and uses the difference $(x_i - \hat{x}_i^j)$ to estimate the contribution of the lower-layer neurons recursively. The contribution re-distribution is performed as follows:

$$R_i^{(l)} = \sum_j \frac{\partial R_j^{(l+1)}}{\partial x_i}\bigg|_{x_i = \hat{x}_i^j} * (x_i - \hat{x}_i^j) \quad (8)$$

where $R_j^{(l+1)}$ is the contribution of the neuron $j$ at the upper layer $(l+1)$, and $R_i^{(l)}$ is the contribution of the neuron $i$ at the lower layer ($l$). It has been shown in Samek et al. (2016) and Montavon et al. (2017) that for NNs with ReLU activations, Deep Taylor leads to similar results to the LRP$_{\alpha=1,\beta=0}$ rule.

**8) Occlusion-1:** This method (Ancona et al., 2018) estimates the attribution of the output to each of the input features $i$ as the difference between the network's prediction when feature $i$ is included in the input and when it is set to zero:

$$R_{i,n} = \hat{F}(\mathbf{x}_n) - \hat{F}(\mathbf{x}_n|x_{i,n} = 0) \tag{9}$$

## 4. Results

In this section, we compare the XAI heatmaps estimated by the considered XAI methods to the ground truth of attribution for $F$ (see step 4 in Fig. 1). The correlation coefficient (see also Ancona et al., 2018; Adebayo et al., 2020) of the estimated heatmap and the ground truth will serve as our metric to assess the performance of the explanation. We highlight however that a perfect agreement between the two (a correlation of 1) shall not be attainable due to $\hat{F}$ being only a close approximation (i.e., not identical) to $F$. We first present results for specific samples in the testing set, to get a qualitative insight on the XAI performance, and then we present more quantitative summary statistics of the performance across all samples from the testing set.

### 4.1. Illustrative comparisons

In Fig. 3, we present the ground truth of attribution for $F$ and the estimated heatmaps from the considered XAI methods (each heatmap is standardized by the corresponding maximum absolute value within the map). This sample corresponds to a response $y_n = 0.0660$, while the NN predicts 0.0802. Based on the ground truth, features that contributed positively to the response $y_n$ occur mainly over the eastern tropical and southern Pacific Ocean, and the southeastern Indian Ocean. Features with negative contribution occur over the Atlantic Ocean and mainly in the tropics.

Based on the method Gradient, the explanation of the NN prediction is not in agreement at all with the ground truth. In the eastern tropical and southern Pacific Ocean, the method returns negative values instead of positive, and over the tropical Atlantic, positive values (instead of negative) are highlighted. The pattern correlation is very small on the order of -0.08, consistent with the above observations. As theoretically expected, this result indicates that the *sensitivity* of the output to the input is not the same (neither numerically nor conceptually) as the *attribution* of the output to the input (see Ancona et al., 2019; Mamalakis et al., 2022). The method Smooth Gradient performs similarly to the method Gradient, with a correlation coefficient on the order of -0.10.

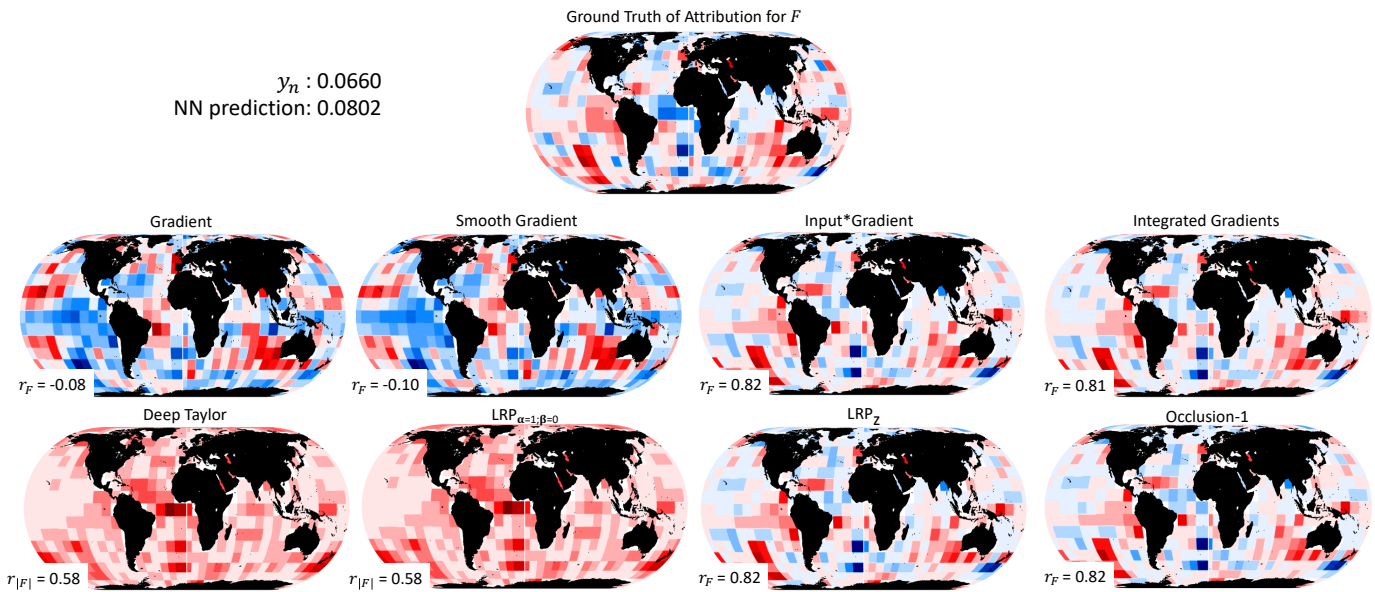

Fig. 3. Performance of different XAI methods for the sample $n = 979{,}476$ in the testing set. The XAI performance is assessed by comparing the estimated heatmaps to the ground truth of attribution for $F$. All heatmaps are standardized with the corresponding maximum (in absolute terms) heatmap value. Red (blue) color corresponds to positive (negative) contribution to (or gradient of) the response/prediction, with darker shading representing higher (in absolute terms) value. For all methods apart from Deep Taylor and LRP$_{\alpha=1,\beta=0}$, the correlation coefficient between the heatmap and the ground truth is also provided. For the methods Deep Taylor and LRP$_{\alpha=1,\beta=0}$ the correlation with the absolute ground truth is given to account for the fact that these two methods do not distinguish between positive and negative attributions (by construction).

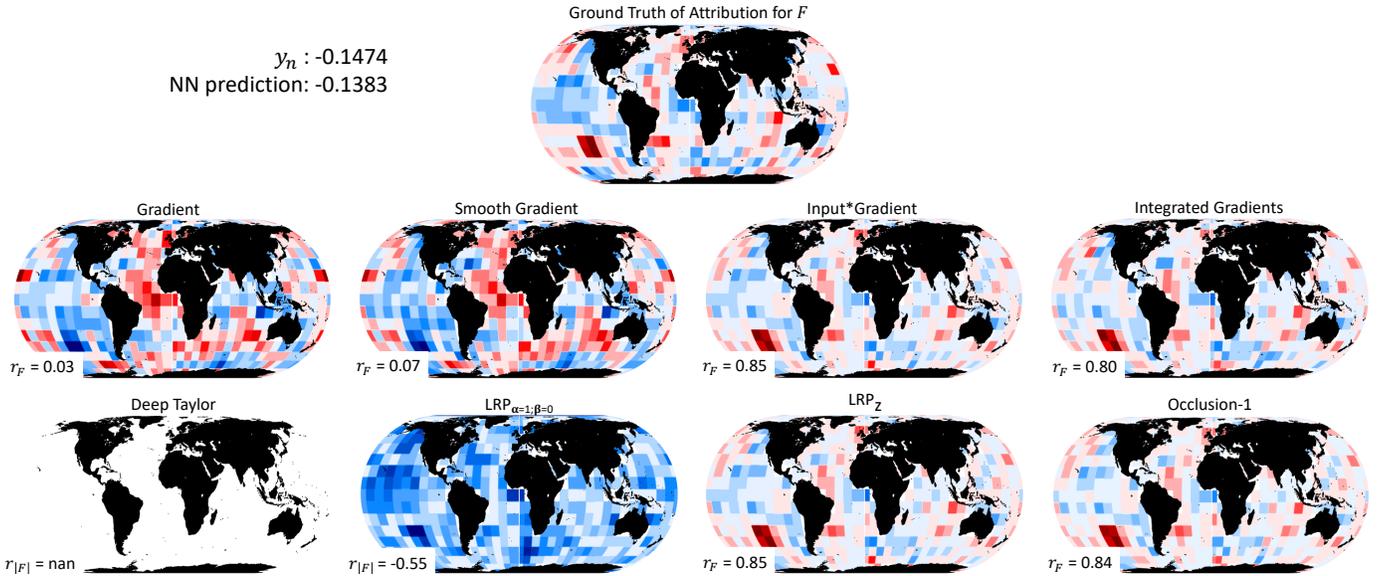

Fig. 4. Same as Fig. 3, but for the sample $n = 995{,}903$ in the testing set.

Methods Input*Gradient and Integrated Gradients perform very similarly, both capturing the ground truth very closely. Indeed, both methods capture the positive patterns over eastern Pacific and the southeastern Indian Oceans, and the negative patterns over the Atlantic Ocean. The pattern correlation with the ground truth for both methods is on the order of 0.8, indicating the very high agreement.

Our results confirm the arguments in Samek et al. (2016) and Montavon et al. (2017), that the Deep Taylor leads to similar results with the LRP$_{\alpha=1,\beta=0}$. Both methods return only non-negative contributions which is not consistent with the ground truth. The inability of the latter methods to distinguish between positive and negative signs can be explained by considering Eq. (7) and setting $\alpha = 1, \beta = 0$ to retrieve the LRP$_{\alpha=1,\beta=0}$ rule. By also noticing that the ratio $\frac{z_{ij}^{+}}{z_j^{+}}$ is by definition a positive number, one can conclude that the contribution of any neuron in the lower layer $R_i^{(l)}$ may only be zero or have the same sign as the contribution of the neuron in the upper layer $R_i^{(l+1)}$, and thus, the sign of the NN prediction is maintained and recursively propagated back to the input layer. Because the NN prediction is positive in Fig. 3, it is expected that LRP$_{\alpha=1,\beta=0}$ (and Deep Taylor) returns non-negative contributions (see also remarks by Kohlbrenner et al., 2020). What is not so intuitive is the fact that the LRP$_{\alpha=1,\beta=0}$ seems to highlight many important features, independent of the sign of their contribution (compare with ground truth). Given that, by construction of Eq. (7), LRP$_{\alpha=1,\beta=0}$ considers only positive preactivations (Bach et al., 2015), one might assume that it will only highlight the features that positively contribute to the prediction.

However, the results in Fig. 3 show that the method highlights the entire Atlantic Ocean with a positive contribution. This is problematic, since the ground truth heatmap clearly indicates that this region is contributing negatively to the response $y_n$ in this example. The issue of LRP$_{\alpha=1,\beta=0}$ in highlighting features independent of whether they are contributing positively or negatively to the prediction has been very recently shown in other applications of XAI as well (Kohlbrenner et al., 2020). Interestingly though, we have established that this issue is not present when one applies the LRP$_{\alpha=1,\beta=0}$ to explain a linear model (not shown). In this case, the LRP$_{\alpha=1,\beta=0}$ returns only the features with positive contribution. This seems to suggest that the issue of mixing positive and negative contributions depends on the complexity of the model that is being explained, and is more likely to occur as the model complexity increases. Finally, we note that to account for the fact that Deep Taylor and LRP$_{\alpha=1,\beta=0}$ do not distinguish between the sign of the attribution, we present their correlation with the absolute ground truth. Both methods correlate on the order of 0.58, which is lower than methods Input*Gradient and Integrated Gradients.

When using the LRP$_z$ and Occlusion-1, the attribution heatmaps very closely capture the ground truth, and they both exhibit a very high pattern correlation on the order of 0.82. The results are very similar to those of the methods Input*Gradient and Integrated Gradients, making these four methods the best performing ones for this example sample. The similarity of these four methods is consistent with the discussion in Anconca et al. (2018) and is based on the fact that all four methods can be mathematically represented as an element-wise product of the input and a modified gradient term (see Table 1 in Anconca et al., 2018). In fact, under specific conditions (i.e., specific network characteristics), some of these methods become exactly equivalent to each other. For example, the methods Input*Gradient and $LRP_z$ are equivalent in cases of NNs with ReLU activation functions, as in our study. We note that when using other activation functions (like Sigmoid or Softplus), $LRP_z$ has empirically been shown to fail and diverge from the other methods (Anconca et al., 2018).

Similar remarks with those based on Fig. 3 can be made based on Fig. 4, which presents the ground truth and the estimated attributions for another example, where the response is negative and equal to $y_n$ = -0.1474. The prediction of the NN for this example is -0.1383. First, methods Gradient and Smooth Gradient significantly differ from the ground truth again, with correlations on the order of 0.03 and 0.07, respectively. Methods Input*Gradient, Integrated Gradients, $LRP_z$ and Occlusion-1 are the best performing ones, all of which strongly correlate with the ground truth (correlation coefficients on the order of 0.8-0.85). The method Deep Taylor does not return any attributions since is defined for only positive predictions (Montavon et al., 2017), a fact that limits its application to classification problems or regression problems with positive predictand variables only. Lastly, in accordance with the remarks from Fig. 3, the attributions from $LRP_{\alpha=1,\beta=0}$ are all non-positive, since the NN prediction for this example is negative. Also, it is again evident that $LRP_{\alpha=1,\beta=0}$ highlights many important features independent of the sign of their true contribution and not only the ones that are positively contributing to the prediction. In general, one should be cautious when using this rule, keeping always in mind that, i) it propagates the sign of the prediction back to the contributions of the input layer and ii) it is likely to mix positive and negative contributions.

### 4.2. Quantitative summarizing statistics

In Fig. 5 we present histograms of the correlation coefficients between different XAI methods and the ground truth for all 100,000 testing samples. In this way one can inspect the performance of each of the XAI methods based on all testing samples and verify the specific remarks that were highlighted above.

First, in panel Fig. 5a, we present results from the same XAI method (i.e., Input*Gradient) but applied to the two different models, the NN and the linear regression model. Thus, any difference in the performance comes solely from how well the corresponding models have captured the true underlying function $F$. The NN more closely approximates the function $F$ since the pattern correlations are systematically higher than the ones for the linear model, consistent with the much better prediction performance of the NN. The average pattern correlation between the attribution of the NN and the ground truth is on the order of 0.8, whereas for the linear model it is on the order of 0.55.

Second, in panel Fig. 5b, we present results for all XAI methods except for LRP, as applied to the NN. Methods Gradient and Smooth Gradient exhibit almost zero average correlation with the ground truth, while methods Input*Gradient, Integrated Gradients and Occlusion-1 perform equally well, exhibiting an average correlation with the ground truth around 0.8.

Last, in panel Fig. 5c, we present results for LRP. The $LRP_z$ rule is seen to be the best performing with very similar performance to the Input*Gradient, Integrated Gradients and Occlusion-1 (as theoretically expected for this model setting; see Ancona et al., 2018). The corresponding average correlation coefficient is on the order of 0.8. Regarding the $LRP_{\alpha=1,\beta=0}$ rule, we present two curves. The first curve (black curve in Fig. 5c) corresponds to correlation with the ground truth after we have set all the negative contributions in the ground truth to zero. The second curve (blue curve) corresponds to correlation with the absolute value of the ground truth. For both curves we multiply the correlation value with -1 when the NN prediction was negative, to account for the fact that the prediction's sign is propagated back to the attributions. Our results show that when correlating with the absolute ground truth (blue curve), the correlations are systematically higher than when correlating with then nonnegative ground truth (black curve). This result verifies that the issue of $LRP_{\alpha=1,\beta=0}$ mixing positive and negative attributions occurs for all the testing samples, further highlighting the need to be cautious when using this rule. Also note that $LRP_{\alpha=1,\beta=0}$ correlates systematically less strongly with the ground truth than $LRP_z$. This suggests that $LRP_{\alpha=1,\beta=0}$ does not estimate the magnitude of the attribution as effectively, most likely because it completely neglects the negative preactivations $z_{ij}^-$ of the forward pass through the network (see Eq. (7)).

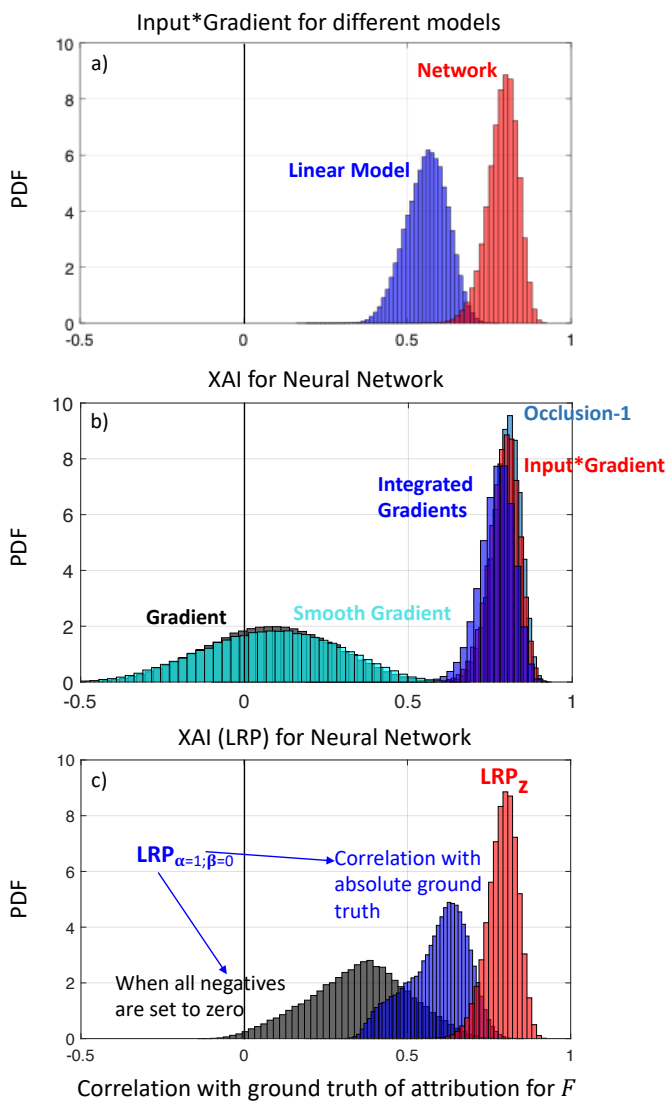

Fig. 5. Summary of the performance of different XAI methods. Histograms of the correlation coefficients between different XAI heatmaps and the ground truth of attribution for 100,000 testing samples.

## 5. Conclusions and future work

The potential for NNs to successfully tackle complex problems in geoscience has become quite evident in recent years. An important requirement for further application of NNs in geoscience is their robust explainability, and newly developed XAI methods show very promising results for this task. However, the assessment of XAI methods often requires the scientist to know what the attribution should look like and is often subjective. Also, applicable attribution benchmark datasets are rarely available, especially for regression problems.

Here, we introduce a new framework to generate synthetic attribution benchmarks to test XAI methods. In our proposed framework, the ground truth of the attribution of the synthetic output to the synthetic input is objectively derivable for any sample. This framework is based on the use of additively separable functions, where the response $Y \in \mathbb{R}$ to the input $\mathbf{X} \in \mathbb{R}^d$ is the sum of local responses. The local responses may have any functional form, while spatial functional dependence can also be enforced. Independent of how complex the local functions might be, the true attribution is always derivable. As an example, we create $10^6$ samples using local piece-wise linear functions and utilize a fully-connected NN to learn the underlying function. Based on the true attribution, we then quantitatively assess the performance of various common XAI methods.

In general, our results suggest that methods Gradient and Smooth Gradient may be suitable for estimating the sensitivity of the output to the input that may offer great insights about the network, but this is not equivalent to the attribution. We also reveal some potential issues in deriving the attribution when using the $LRP_{\alpha=1,\beta=0}$ rule. For the specific setup used here, the methods Input*Gradient, Integrated Gradients, Occlusion-1, and the $LRP_z$ rule all very closely capture the true function $F$ and are the best performing XAI methods considered here.

In summary, in this study we demonstrated the benefits of attribution benchmarks for the identification of possible systematic pitfalls of XAI, and introduced a framework to create such benchmarks with emphasis on geoscience. Clearly, this is only the beginning of a larger research effort. In the future, we plan to extend this work to assess a larger range of XAI methods, using different deep learning models (convolutional NNs, recurrent NNs, etc.) and to derive other forms of nonlinear local functions encountered more frequently in the geosciences (e.g., based on ordinary differential equations). We believe that a common use and engagement of such attribution benchmarks by the geoscientific community can lead to a more cautious and accurate application of XAI methods to physical problems. Such efforts will increase model trust and facilitate scientific discovery.

### Funding Statement

This work was supported in part by the National Science Foundation under Grant No. OAC-1934668. I. E.-U. also acknowledges the support by the National Science Foundation under Grant No. ICER-2019758.

### Competing interests

The authors report no competing interests.

### Data availability

The data that support the findings will be made publicly available upon publication. The code has been made publicly available at: https://github.com/amamalak/Neural-Network-Attribution-Benchmark-for-Regression

## Author contributions

Conceptualization, A.M., I.E.-U. and E.A.B; Methodology, A.M., I.E.-U. and E.A.B; Formal Analysis, A.M.; Data Curation, A.M.; Writing—Original Draft, A.M., Writing—Review and Editing, A.M., I.E.-U. and E.A.B; Supervision, I.E.-U. and E.A.B.; Funding Acquisition, I.E.-U. and E.A.B.

## References

Adebayo, J., *et al.*, "Sanity checks for saliency maps," arXiv preprint, https://arxiv.org/abs/1810.03292, 2020.

Agapiou, A., "Remote sensing in a petabyte-scale: satellite data and heritage Earth Engine© applications, *International Journal of Digital Earth*, vol. 10, no. 1, pp. 85-102, 2017.

Ancona, M., *et al.*, "Gradient-based attribution methods," in *Explainable AI: Interpreting, Explaining and Visualizing Deep Learning*, pp. 169-191, Springer, 2019.

Ancona, M., *et al.,* "Towards better understanding of gradient-based attribution methods for deep neural networks," arXiv preprint, https://arxiv.org/abs/1711.06104, 2018.

Bach, S., *et al.,* "On pixel-wise explanations for non-linear classifier decisions by layer-wise relevance propagation," *PLoS One*, vol. 10, no. 7, e0130140, 2015.

Barnes, E.A., *et al.,* "Viewing forced climate patterns through an AI Lens," *Geophysical Research Letters*, vol. 46, pp. 13389-13398, 2019.

Barnes, E.A., *et al.,* "Indicator patterns of forced changed learned by an artificial neural network," *Journal of Advances in Modeling Earth Systems*, vol. 12, e2020MS002195, 2020.

Bergen, K.J., *et al.,* "Machine learning for data-driven discovery in solid Earth geoscience," *Science*, vol. 363, no. 6433, eaau0323, 2019.

Buhrmester, V., D. Münch, M. Arens, "Analysis of explainers of black box deep neural networks for computer vision: A survey," arXiv preprint, https://arxiv.org/abs/1911.12116, 2019.

Das, A., P. Rad, "Opportunities and challenges in explainable artificial intelligence (XAI): A survey," arXiv preprint, https://arxiv.org/abs/2006.11371, 2020.

DelSole, T., A. Banerjee, "Statistical seasonal prediction based on regularized regression," *Journal of Climate*, vol. 30, no. 4, pp. 1345-1361, 2017.

Dombrowski, A.-K., *et al.,* "Towards robust explanations for deep neural networks," arXiv preprint, https://arxiv.org/abs/2012.10425, 2020.

Ebert-Uphoff, I., K. Hilburn, "Evaluation, tuning, and interpretation of neural networks for working with images in meteorological applications," *Bulletin of the American Meteorological Society*, vol. 101, no. 12, pp. E2149-E2170, 2020.

Feng, J., W. Chen, Y. Li, "Asymmetry of the winter extra-tropical teleconnections in the Northern Hemisphere associated with two types of ENSO," *Clim. Dyn.*, vol. 48, pp. 2135-2151, 2017.

Guo, H., "Big Earth data: A new frontier in Earth and information sciences," *Big Earth Data*, vol. 1, no. 1-2, pp. 4-20, 2017.

Ham, Y.G., J. H. Kim, J. J. Luo, "Deep learning for multi-year ENSO forecasts," *Nature*, vol. 573, pp. 568-572, 2019.

Karpatne, A., *et al.,* "Machine learning for the Geosciences: Challenges and opportunities," *IEEE Trans. Knowledge and Data Engin.*, vol. 31, no. 8, pp. 1544-1554, 2018.

Kohlbrenner, M., *et al.*, "Towards best practice in explaining neural network decisions with LRP," 2020 International Joint Conference on Neural Networks (IJCNN), Glasgow, UK, pp. 1-7, 2020.

Kindermans, P.-J., *et al.*, "Learning how to explain neural networks: PatternNet and PatternAttribution," arXiv preprint, https://arxiv.org/abs/1705.05598, 2017a.

Kindermans, P.-J., *et al.*, "The (un)reliability of saliency methods," arXiv preprint, https://arxiv.org/abs/1711.00867, 2017b.

Lapuschkin, S., *et al.,* "Unmasking clever Hans predictors and assessing what machines really learn," *Nature Communications*, vol. 10, no. 1096, 2019.

Lary, D.J., *et al.,* "Machine learning in geosciences and remote sensing," *Geoscience Frontiers*, vol. 7, no. 1, pp. 3-10, 2016.

Leavitt, M.L., A.S. Morcos, "Towards falsifiable interpretability research," arXiv preprint, https://arxiv.org/abs/2010.12016, 2020.

LeCun, Y., Y. Bengio, G. Hinton, "Deep learning," *Nature*, vol. 521, pp. 436-444, 2015.

Mamalakis, A., E. A. Barnes and I. Ebert-Uphoff, "Investigating the fidelity of explainable artificial intelligence methods for applications of convolutional neural networks in geoscience," arXiv preprint, https://arxiv.org/abs/2202.03407, 2022.

McGovern, A., *et al.,* "Making the black box more transparent: Understanding the physical implications of machine learning," *Bulletin of the American Meteorological Society*, vol. 100, no. 11, pp. 2175-2199, 2019.

Montavon, G., *et al.,* "Explaining nonlinear classification decisions with deep Taylor decomposition," *Pattern Recognition*, vol. 65, pp. 211-222, 2017.

LeCun, Y., *et al.,* "Gradient-based learning applied to document recognition," *Proceedings of the IEEE*, vol. 86, no. 11, pp. 2278-2324, 1998.

Overpeck, J.T., *et al.,* "Climate data challenges in the 21$^{st}$ century," *Science*, vol. 331, no. 6018, pp. 700-702, 2011.

Petsiuk, V., A. Das, K. Saenko, "RISE: Randomized input sampling for explanation of black-boc models," arXiv preprint, https://arxiv.org/abs/1806.07421, 2018.

Qi, Z., S. Khorram, L. Fuxin, "Visualizing deep networks by optimizing with integrated gradients," arXiv preprint, https://arxiv.org/abs/1905.00954, 2020.

Reichstein, M., *et al.,* "Deep learning and process understanding for data-driven Earth system science," *Nature*, vol. 566, pp. 195-204, 2019.

Reinsel, D., J. Gantz, J. Rydning, "The digitization of the world: from edge to core" IDC, Framingham, MA, USA, White Paper Doc# US44413318, 2018, p. 28. [Online]. Available: https://www.seagate. com/files/www-content/our-story/trends/files/idc-seagate-dataagewhitepaper.pdf

Rolnick, D., *et al.*, "Tackling climate change with machine learning, arXiv preprint, https://arxiv.org/abs/1906.05433, 2019.

Rudin, C., "Stop explaining black box machine learning models for high stakes decisions and use interpretable models instead," *Nature Machine Learning*, vol 1, pp. 206-215, 2019.

Russakovsky, O., *et al.,* "ImageNet large scale visual recognition challenge," arXiv preprint, https://arxiv.org/abs/1409.0575, 2015.

Samek, W., *et al.*, "Evaluating the visualization of what a deep neural network has learned," *IEEE Transactions on Neural Networks and Learning Systems*, vol. 28, no. 11, 2017.

Samek, W., *et al.,* "Interpreting the predictions of complex ML models by layer-wise relevance propagation," arXiv preprint, https://arxiv.org/abs/1611.08191, 2016.

Shen, C., "A transdisciplinary review of deep learning research and its relevance for water resources scientists," *Water Resources Research*, vol. 54, pp. 8558-8593, 2018.

Shrikumar, A., *et al.,* "Not just a black box: Learning important features through propagating activation differences," arXiv preprint, https://arxiv.org/abs/1605.01713, 2016.

Shrikumar, A., P. Greenside, A. Kundaje, "Learning important features through propagating activation differences," arXiv preprint, https://arxiv.org/abs/1704.02685, 2017.

Sit, M., *et al.,* "A comprehensive review of deep learning applications in hydrology and water resources," arXiv preprint, https://arxiv.org/abs/2007.12269, 2020.

Smilkov, D., *et al.*, "SmoothGrad: removing noise by adding noise," arXiv preprint, https://arxiv.org/abs/1706.03825, 2017.

Springenberg, J.T., *et al.,* "Striving for simplicity: The all convolutional net," arXiv preprint, https://arxiv.org/abs/1412.6806, 2015.

Stevens, A., *et al.,* "Graph-guided regularized regression of Pacific Ocean climate variables to increase predictive skill of southwestern US winter precipitation," *Journal of Climate*, vol. 34, no. 2, pp. 737-754, 2021.

Sundararajan, M., A. Taly, Q. Yan, "Axiomatic attribution for deep networks," arXiv preprint, https://arxiv.org/abs/1703.01365, 2017.

Toms, B.A., E. A. Barnes, I. Ebert-Uphoff, "Physically interpretable neural networks for the geosciences: Applications to Earth system variability," *Journal of Advances in Modeling Earth Systems*, vol. 12, e2019MS002002, 2020.

Tjoa, E., C. Guan, "A survey on explainable artificial intelligence (XAI): Towards medical XAI," arXiv preprint, arXiv:1907.07374, 2019.

Zeiler, M.D., R. Fergus, "Visualizing and understanding convolutional networks," arXiv preprint, https://arxiv.org/abs/1311.2901, 2013.

Zhang, T., J. Perlwitz, M. P. Hoerling, "What is responsible for the strong observed asymmetry in teleconnections between El Niño and La Niña?," *Geophysical Research Letters*, vol. 41, pp. 1019-1025, 2014.